\documentclass[sigconf,nonacm]{acmart}

\AtBeginDocument{%
  }

\usepackage{tikz}
\usetikzlibrary{positioning,arrows.meta,fit,shadows,calc}
\usepackage{algorithm}
\usepackage{algpseudocode}
\usepackage{graphicx}
\usepackage{booktabs}
\usepackage{array}
\usepackage{float}
\usepackage{enumitem}
\newcommand{\GPTCatalog}{GPT\allowbreak{}+\allowbreak{}Catalog}
\newcommand{\GeminiCatalog}{Gemini\allowbreak{}+\allowbreak{}Catalog}

\begin{document}

\title[Reliable Conversational Recommendation for Agentic E-Commerce]{
MACS: A Hybrid Multi-Agent Framework for Reliable Conversational E-Commerce Recommendation
}

\author{Juli Huang}
\affiliation{%
  \institution{Stanford University}
  \department{Department of Computer Science}
  \city{Stanford}\state{CA}\country{USA}}
\email{julih@stanford.edu}

\author{Hannah Clay}
\affiliation{%
  \institution{Stanford University}
  \department{Department of Computer Science}
  \city{Stanford}\state{CA}\country{USA}}
\email{hclay116@stanford.edu}

\author{Thomas Sarda}
\authornote{Now at Google DeepMind.}
\affiliation{%
  \institution{Stanford University}
  \department{Department of Computer Science}
  \city{Stanford}\state{CA}\country{USA}}
\email{tsarda@stanford.edu}

\author{Sajjad Beygi}
\affiliation{%
  \institution{University of Southern California}
  \city{Los Angeles}\state{CA}\country{USA}}
\affiliation{%
  \institution{Amazon}
  \country{USA}}
\email{beygi.ee@gmail.com}

\author{Negin Golrezaei}
\affiliation{%
  \institution{Massachusetts Institute of Technology}
  \department{MIT Sloan School of Management}
  \city{Cambridge}\state{MA}\country{USA}}
\email{golrezae@mit.edu}

\author{Amin Saberi}
\affiliation{%
  \institution{Stanford University}
  \department{Department of Management Science \& Engineering}
  \city{Stanford}\state{CA}\country{USA}}
\email{saberi@stanford.edu}

\renewcommand{\shortauthors}{Huang et al.}

\begin{abstract}
Conversational recommendation for e-commerce is increasingly mediated by large language models (LLMs), yet many real-world deployments operate under a stricter requirement: recommendations must be drawn only from a merchant's fixed catalog, without web search or unsupported product claims. In this setting, the main challenge is reliability under hard constraints: the system must satisfy user requirements, remain grounded in available inventory, and preserve preferences across multiple conversational turns. We present MACS (Multi-Agent Commerce System), a hybrid multi-agent framework for reliable conversational recommendation in fixed-catalog settings. MACS uses LLMs for language-facing tasks such as interpreting user requests, eliciting preferences, and generating responses, while correctness-critical operations, including product retrieval, hard-constraint filtering, brand exclusion, and progressive relaxation, are executed deterministically by the merchant agent. A session-persistent preference layer tracks constraints across turns, enabling consistent handling of budget overwrites and exclusion reversals. On a 140-query single-turn benchmark, MACS achieves the highest pass rate (87.1\%) and perfect brand compliance (1.000). On a 10-scenario multi-turn benchmark, MACS achieves the strongest macro Pass@5 (72\% vs.\ 56\% \GPTCatalog{} / 52\% \GeminiCatalog{}) with zero constraint drift. The advantage is sharpest on exclusion reversal (100\% vs.\ 20\% / 0\%) and constraint accumulation (100\% vs.\ 60\% / 40\%). Mean judged response quality is similar across systems (0.751 vs.\ 0.736). These results suggest that hybrid architectures combining deterministic constraint enforcement with session-persistent preference tracking provide stronger reliability-oriented performance than catalog-bound prompt-only baselines in the fixed-catalog merchant setting.

\end{abstract}



\maketitle


\section{Introduction}
\label{sec:introduction}

Conversational recommender systems increasingly rely on large language models
(LLMs) to interact with users in natural language, elicit preferences, and
generate recommendations. A particularly important and underexplored deployment
setting is \emph{limited-catalog shopping}: a merchant agent using its
own fixed product database without web search to enforce the recommendation of
verified inventory while respecting user constraints. This setting gives merchants control over product data and avoids the latency and hallucination risks of web-augmented search. However, it places strict reliability demands on the recommender. LLMs, despite their fluency, remain unreliable in this setting: they may hallucinate product specifications not present in the catalog, forget earlier brand exclusions or other user preferences across conversational turns, or apply budget filters inconsistently. Therefore, a recommendation system that returns out-of-catalog products or violates a user's stated constraints fails a core reliability requirement.

\textbf{MACS (Multi-Agent Commerce System)} is a hybrid agent-to-agent e-commerce framework for reliable conversational recommendation in fixed-catalog settings. MACS is composed of two cooperating agents: a \emph{shopping agent} that handles natural-language input, intent routing, and response generation; and a \emph{merchant agent} that enforces hard constraints deterministically and retrieves products exclusively from the fixed catalog. The shopping agent communicates with the merchant agent via structured commerce protocols, specifically Universal Commerce Protocol (UCP)~\cite{Google2026UCP} or Agentic Commerce Protocol (ACP)~\cite{OpenAI2025ACP}. Because the shopping agent cannot retrieve products directly, catalog grounding is enforced at the architectural level rather than through prompt instructions.

Three core contributions are made as follows:
\begin{itemize}[leftmargin=*]
    \item \textbf{Architecture.} A hybrid shopping-agent to merchant-agent architecture
    separates LLM-based natural-language interaction from deterministic constraint
    enforcement, enabling reliable product recommendation over a fixed,
    merchant-owned catalog without web search.

    \item \textbf{Session state.} Session-persistent preference state management
    preserves user constraints across turns, supports updates such as budget
    overwrites and exclusion reversals, and enables consistent multi-turn
    recommendation behavior.

    \item \textbf{Evaluation framework.} A hybrid evaluation methodology separates deterministic
    constraint correctness from response quality, enabling faithful assessment
    of conversational recommenders in both single-turn and multi-turn
    limited-catalog settings.
\end{itemize}

\section{Related Work}
MACS sits at the intersection of three active research areas:
conversational recommender systems, knowledge-graph-augmented
retrieval, and LLM reliability under hard constraints. Prior work
in each area motivates a component of the MACS design. However, none
addresses the combination of deterministic constraint enforcement,
session-persistent preference state, and fixed-catalog grounding
that defines limited-catalog commerce-agent settings.

\paragraph{Conversational Recommender Systems.}
CRSs elicit user preferences through multi-turn interaction~\cite{Jannach2021Survey},
ranging from bandit-based question selection~\cite{Christakopoulou2016} and RL-based dialogue policy learning~\cite{Sun2018}
to LLM-integrated pipelines for preference understanding and response
generation~\cite{Friedman2023,Feng2023LLMCRS,Liu2023CRSLLM}.
Retrieval augmentation~\cite{Yang2024ReFICR}, cross-session
memory~\cite{Xi2024MemoCRS}, knowledge-enhanced sequential modeling~\cite{zou2024knowledge},
and behavioral evaluation~\cite{Yang2024BehaviorAlignment}
have further extended the paradigm.
Kostric et al.~\cite{Kostric2024ClarifyCRS} show that generating
usage-oriented clarifying questions yields richer preference signals in
multi-turn CRS, underscoring the need for explicit cross-turn
preference-tracking mechanisms.
MACS addresses a gap these works leave open: deterministic enforcement of
hard user constraints (excluded brands, budget ceilings, specification minimums)
across turns, with safeguards to combat LLM unreliability.

\paragraph{Knowledge Graphs and LLM Agents for Shopping.}
Knowledge-aware recommenders (CKE~\cite{Zhang2016CKE}, KGCN~\cite{Wang2019KGCN},
KGAT~\cite{Wang2019KGAT}) incorporate a graph structure into learned representations;
MACS uses a knowledge graph operationally for structural queries (substitutes-(products that serve the same purpose at a similar price point),
compatibility) via explicit Cypher queries over a knowledge graph, so the retrieval path can be directly inspected. By contrast, learned embeddings produce similarity scores that are opaque to inspection.
Shopping-agent benchmarks (WebArena~\cite{Zhou2023WebArena},
AgentBench~\cite{Liu2023AgentBench}, ShoppingBench~\cite{Wang2025ShoppingBench},
$\tau$-Bench~\cite{Yao2024TauBench})
study the \emph{shopping-agent} side; MACS studies both the shopping-agent and complementary
\emph{merchant-agent} design that provides structured, constraint-enforcing catalog access.

\paragraph{LLM Reliability and Preference Elicitation.}
LLM faithfulness under hard constraints (e.g., brand exclusions, budget ceilings,
specification minimums) is a known open problem.
Chain-of-thought prompting reduces but does not eliminate constraint violations in multi-turn settings~\cite{Jannach2021Survey,Feng2023LLMCRS}. Preference elicitation approaches such as EAR~\cite{Lei2020EAR} and UniCRS~\cite{Wang2022UniCRS} ask clarifying questions to narrow item sets before retrieval, but treat constraint enforcement as an LLM inference task rather than a deterministic predicate~\cite{Christakopoulou2016,Sun2018}. MACS decouples these two responsibilities: preference elicitation remains an LLM task, while constraint enforcement is delegated to SQL predicates persisted across turns in a typed session-state layer. As a result, constraint satisfaction does not depend on the quality of the LLM’s generation. Prompt-based approaches cannot guarantee this, because they still rely on the model itself to enforce constraints.


\section{Problem Statement and System Overview}
\label{sec:problem}
We formalize limited-catalog conversational recommendation as a constrained multi-turn retrieval problem over a fixed catalog:
\paragraph{Problem definition.}
A merchant catalog $\mathcal{C}$ contains products with structured attributes
(for example, in the electronics domain: price, brand, RAM, storage, and display size). Over a multi-turn conversation,
the system must extract user constraints, retrieve $R_t \subseteq \mathcal{C}$
satisfying all accumulated constraints at turn $t$, and generate factually
grounded responses. It also handles add-to-cart, checkout, and service queries
(return policy, shipping, warranty) through the same session.

Three reliability requirements follow from the limited-catalog setting:
catalog grounding (every returned product exists at a verified price),
constraint correctness (hard requirements enforced deterministically, not
inferred), and cross-turn persistence (constraints active until explicitly
revised, never silently dropped).

\paragraph{System overview.}
MACS separates a \emph{shopping agent} (LLM-based preference elicitation,
intent routing, response generation) from a \emph{merchant agent} (SQL
constraint filtering, knowledge graph traversal, session-persistent state).
The shopping agent never retrieves products directly; all calls go through
the merchant agent's structured API, so the LLM cannot hallucinate catalog
contents. Figure~\ref{fig:arch} illustrates the two agents and their
interaction diagram. Sections~\ref{sec:shopping-agent}
and~\ref{sec:merchant-agent} describe each component.


\section{The Shopping Agent}
\label{sec:shopping-agent}

On each turn, the shopping agent interprets the user’s message, determines the action to take (for example, new search, refinement, comparison, add-to-cart, checkout, or FAQ), extracts structured preference slots, and requests catalog results from the merchant agent before generating a response. All correctness-critical operations, including product retrieval, hard-constraint enforcement, exclusion of disallowed brands or attributes, and progressive relaxation of lower-priority constraints when necessary, are handled by the merchant agent
(Section~\ref{sec:merchant-agent}).

\subsection{Pipeline: Rewriting, Routing, and Extraction}

Each turn passes through query rewriting, intent routing, structured extraction,
and catalog-grounded execution.
\emph{Query rewriting} injects structured hints before any LLM call
(e.g., ``for my son'' $\to$ \texttt{[use\_case:~school]};
``gaming Chromebook'' $\to$ a ChromeOS-compatibility note).
\emph{Intent routing} intercepts common intents (compare, refine, add-to-cart)
with deterministic keyword rules, while ambiguous requests are handled by an
LLM router.
\emph{Structured extraction} converts the user's message into typed slot-value pairs, such as budget, excluded brands, and required specifications, and resolves follow-up references across turns, including requests like ``show me something cheaper'' or ``add the second one to the cart.''

\subsection{Session State and Multi-Turn Management}

The session state is a typed slot dictionary with three key properties:
\begin{itemize}[leftmargin=*]
\item \emph{Cross-turn accumulation and ordinal resolution.}
  Slot values persist across turns
  unless explicitly revised. References such as ``add the second one'' are
  resolved against the most recent recommendation context. Brand exclusions
  are enforced at the SQL level and in a post-retrieval title filter, so no
  excluded brand appears regardless of database condition labels.
\item \emph{Preference pivot detection.}
  Updated constraints override earlier ones when appropriate: a new budget
  ceiling replaces the previous value, and a statement such as ``actually HP
  is fine'' reverses an earlier brand exclusion.
\item \emph{Underspecification handling.}
  When the query contains no hard constraint (brand exclusion, price ceiling,
  or specification minimum) and the message is fewer than four words, the agent
  asks a targeted follow-up question rather than guessing. When at least one
  hard constraint is present with a sufficiently specific message, it proceeds
  directly to search.
\end{itemize}

The recommendation output presents products that satisfy the current constraints, explains why they were selected, and provides follow-up options for refinement, comparison, and further catalog browsing.

\section{The Merchant Agent}
\label{sec:merchant-agent}

The merchant agent enforces hard constraints deterministically (SQL filtering,
brand exclusion, progressive relaxation) and resolves complex structural
queries via knowledge graph traversal. Operating exclusively over the
merchant's catalog, it guarantees that every product exists at a verified
price.

\begin{figure*}[t]
\centering
\resizebox{\textwidth}{!}{%
\begin{tikzpicture}[
    box/.style={
        rectangle,
        draw,
        thick,
        rounded corners=4pt,
        align=center,
        minimum height=0.95cm,
        font=\small
    },
    arrow/.style={
        -{Latex[length=2.5mm]},
        thick,
        >=latex
    },
    title/.style={font=\normalsize\bfseries, anchor=west, inner sep=2pt},
    node distance=0.55cm and 0.7cm
]

\def\colA{2cm}
\def\colB{2.5cm}
\def\colC{2.6cm}
\def\colD{2.2cm}
\def\colE{2.2cm}
\def\colF{2cm}
\def\colGap{1cm}

\node[title] (f1title) at (0,0)
  {\textbf{Flow 1: Search} --- ``looking for a gaming laptop''};

\node[box, fill=gray!10, below=0.9cm of f1title.north west, anchor=north west,
      minimum width=\colA] (u1) {
  \textbf{User}\\``gaming laptop''};

\node[box, fill=blue!12, right=\colGap of u1.east, anchor=west,
      minimum width=\colB] (ai1a) {
  \textbf{AI Shopping Agent}\\Parse request\\Generate UCP request};

\node[box, fill=green!12, right=\colGap of ai1a.east, anchor=west,
      minimum width=\colC] (ma1a) {
  \textbf{Merchant Agent}\\Parse UCP request\\Use KG $\to$ items\\Prices, shipping};

\node[box, fill=green!12, right=\colGap of ma1a.east, anchor=west,
      minimum width=\colD] (ma1b) {Create UCP\\response};

\node[box, fill=blue!12, right=\colGap of ma1b.east, anchor=west,
      minimum width=\colE] (ai1b) {
  \textbf{AI Shopping Agent}\\Receive response\\Present to user};

\node[box, fill=gray!10, right=\colGap of ai1b.east, anchor=west,
      minimum width=\colF] (u1out) {\textbf{User}\\Sees results};

\draw[arrow] (u1) -- (ai1a);
\draw[arrow] (ai1a) -- (ma1a);
\draw[arrow] (ma1a) -- (ma1b);
\draw[arrow] (ma1b) -- (ai1b);
\draw[arrow] (ai1b) -- (u1out);

\node[title, anchor=west] (f2title)
  at ([yshift=-0.6cm]f1title.west |- u1.south)
  {\textbf{Flow 2: Add to Cart} --- ``add this to cart''};

\node[box, fill=gray!10, below=0.9cm of f2title.north west, anchor=north west,
      minimum width=\colA] (u2) {
  \textbf{User}\\Select product\\Add to cart};

\node[box, fill=blue!12, right=\colGap of u2.east, anchor=west,
      minimum width=\colB] (ai2a) {
  \textbf{AI Shopping Agent}\\Parse query\\Generate UCP request};

\node[box, fill=green!12, right=\colGap of ai2a.east, anchor=west,
      minimum width=\colC] (ma2a) {
  \textbf{Merchant Agent}\\Parse UCP request\\Add to cart\\Validate inventory};

\node[box, fill=green!12, right=\colGap of ma2a.east, anchor=west,
      minimum width=\colD] (ma2b) {Create UCP\\response};

\node[box, fill=blue!12, right=\colGap of ma2b.east, anchor=west,
      minimum width=\colE] (ai2b) {
  \textbf{AI Shopping Agent}\\Receive response\\Present to user};

\node[box, fill=gray!10, right=\colGap of ai2b.east, anchor=west,
      minimum width=\colF] (u2out) {\textbf{User}\\Cart updated};

\draw[arrow] (u2) -- (ai2a);
\draw[arrow] (ai2a) -- (ma2a);
\draw[arrow] (ma2a) -- (ma2b);
\draw[arrow] (ma2b) -- (ai2b);
\draw[arrow] (ai2b) -- (u2out);

\end{tikzpicture}%
}
\caption{MACS interaction flows. The AI shopping agent handles natural-language
  interaction and protocol generation; the merchant agent enforces hard constraints,
  validates inventory, and returns structured responses. Flow~1 (Search): the user's
  query is parsed, constraint-filtered via the knowledge graph and SQL store, and
  results are returned. Flow~2 (Add to Cart): the merchant agent validates inventory
  before confirming. All constraint enforcement occurs exclusively in the merchant agent.}
\Description{Two interaction flow diagrams. Flow 1 (Search): User to AI Shopping Agent
  (parse request, generate UCP) to Merchant Agent (parse UCP, use KG, prices) to
  Create UCP response to AI Shopping Agent (present to user) to User (sees results).
  Flow 2 (Add to Cart): User to AI Shopping Agent to Merchant Agent (add to cart,
  validate inventory) to Create UCP response to AI Shopping Agent to User (cart updated).}
\label{fig:arch}
\end{figure*}

\subsection{Protocol and Data Layer}

MACS exposes three protocol families through a shared backend interface, so adding a new protocol requires only a request adapter and response formatter.
\textbf{Model Context Protocol (MCP)}~\cite{Anthropic2024MCP} provides four typed tool calls
(\texttt{search\_products}, \texttt{get\_product}, \texttt{add\_to\_cart},
\texttt{checkout}) with observable response envelopes.
\textbf{Universal Commerce Protocol (UCP)}~\cite{Google2026UCP} follows Google's schema for Gemini-based agents.
\textbf{Agent Commerce Protocol (ACP)}~\cite{OpenAI2025ACP} adds a checkout session lifecycle and webhooks.
A configuration variable selects UCP or ACP at deployment time.
\subsection{Data Layer}

Three data stores serve distinct roles.

\textbf{The relational product store} holds the authoritative product catalog
(21,000+ items; 1,490+ laptops with 12 structured specification fields per product:
CPU, RAM, GPU, storage type, display size, battery life, weight, OS, price,
brand, review count, and rating).
Constraint filtering, brand exclusion, and progressive relaxation all execute as SQL queries against this store.

\textbf{The knowledge graph} encodes product relationships across
2,400 nodes and 8,500 edges in two actively-queried relationship types:
\textsc{Similar\_To} (same use-case substitutes at a similar price) and
\textsc{Compatible\_With} (accessory compatibility).
Complex intent queries (``show me similar items,'' ``find a compatible
accessory'') are resolved via Cypher traversal in $\sim$17~ms with no LLM call.
The graph degrades gracefully to SQL-based search when unavailable,
ensuring no hard dependency on the graph layer.

\textbf{The cache and session layer} caches search results, product summaries,
price data, and LLM-generated narratives using per-entry Time-To-Live (TTL)
expirations (30~secs-30~min depending on data freshness requirements).
This caching layer achieves $\sim$12$\times$ latency reduction on cache hits
(36~ms vs.\ 446~ms cold read, measured over $n = 10$ runs).


\subsection{Hard Constraint Filtering}

Extracted slot values map directly to SQL \texttt{WHERE} clauses. Brand exclusions are enforced both at the SQL level and in a post-retrieval title filter. This dual enforcement is necessary because some entries in our catalog store condition metadata (e.g., ``New'', ``Recertified'') in the brand field rather than the manufacturer name.

\subsection{Progressive Relaxation}

When hard constraints produce fewer than three results, the system
applies a sequence of relaxation steps ordered by constraint importance.
At each step, the least-critical optional specification constraint is dropped first. Price ceilings and brand exclusions, however, are never relaxed.
The user always sees real, purchasable products; the system never returns an
empty result set or fabricated alternatives.
When relaxation occurs, the response explicitly discloses which constraints
were loosened, so the user is not misled about what was actually enforced.

\begin{figure}[t]
  \centering
  \includegraphics[width=\linewidth]{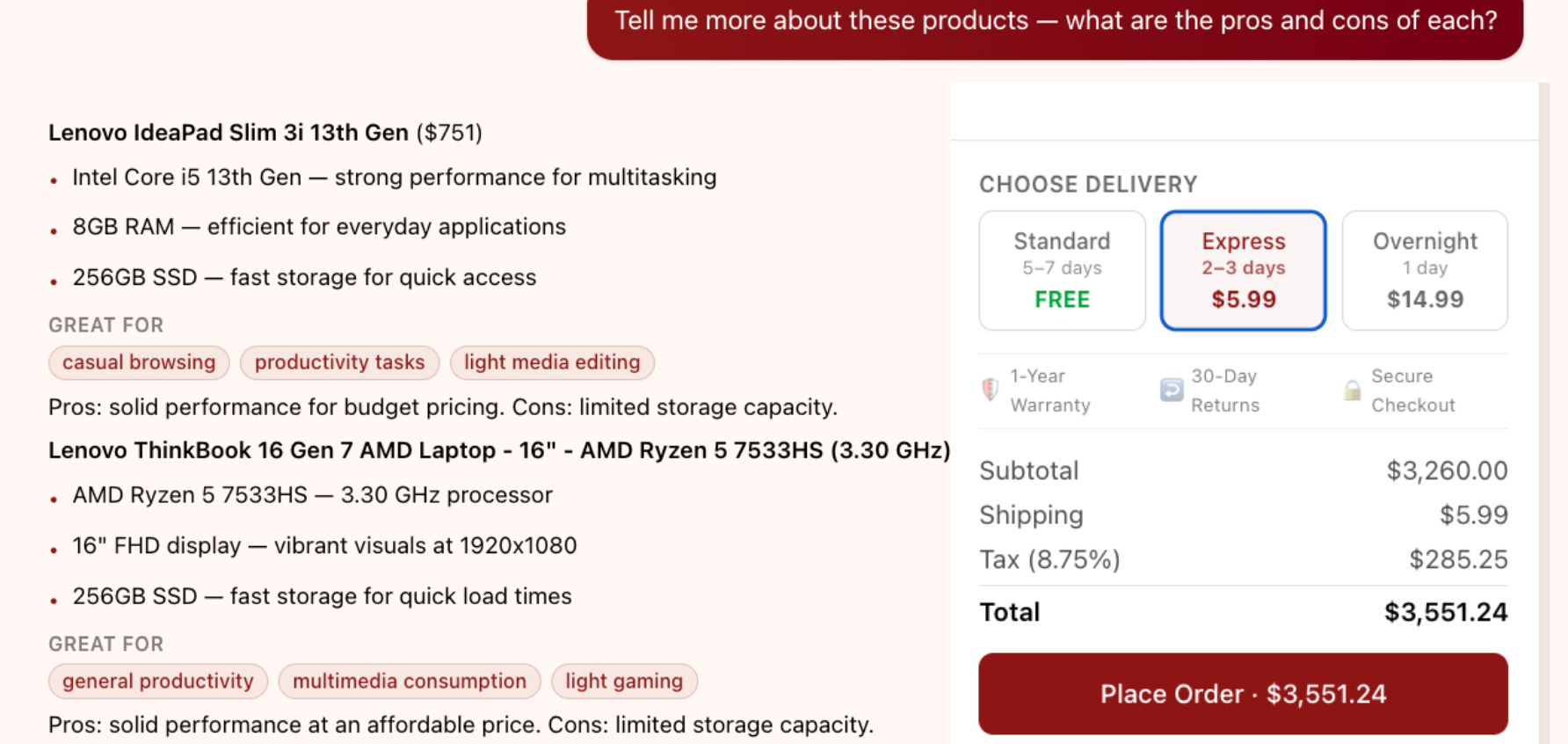}
  \caption{MACS comparison and checkout view. Pros/cons are derived from
    structured catalog attributes, and delivery options and pricing are retrieved
    directly from the merchant agent's relational store, not generated by the LLM.}
  \Description{Side-by-side product comparison showing Lenovo IdeaPad vs.\
    ThinkBook with pros/cons bullet points, delivery options (Standard, Express,
    Overnight), and itemized pricing breakdown.}
  \label{fig:comparison}
\end{figure}

Figure~\ref{fig:comparison} shows the comparison and checkout view, in which
side-by-side pros/cons and delivery options are resolved from the relational
store and presented without LLM-generated fabrication.

\subsection{Best-Value Scoring}

The best-value pick uses a weighted scoring function tuned to the detected
use case:

\begin{equation}
\text{score}(p) = w_{\text{price}} \cdot r_p +
                  w_{\text{rating}} \cdot \hat{r} +
                  \min\!\bigl(n/200,\;c_{\text{vol}}\bigr) +
                  w_{\text{spec}} \cdot \phi(p)
\end{equation}

where $p$ denotes a product; $r_p \in [0,1]$ is the normalized price score
computed as $1 - ({\rm price}_p - {\rm price}_{\min}) / ({\rm price}_{\max} -
{\rm price}_{\min})$, so lower-priced products receive higher scores within
the retrieved set; $\hat{r}$ is the star rating; $n$ is review count;
$c_{\text{vol}}$ is a use-case-specific cap on the review-volume confidence
boost (so that a single viral product cannot dominate by review count alone);
and $\phi(p) \in [0,1]$ is a use-case specification score: a deterministic weighted
sum over structured specification fields (e.g., GPU tier for gaming, RAM tier for ML/video editing, battery life for
student use). Unlike a learned similarity score, $\phi(p)$ is computed directly from catalog attributes.
Table~\ref{tab:weights} reports the full weight vectors.

\begin{table}[h]
\centering
\caption{Best-value scoring weights and $\phi(p)$ spec-score components per use case ($c_{\text{vol}}=0.05$ for all).}
\label{tab:weights}
\resizebox{\columnwidth}{!}{%
\small\begin{tabular}{lcccp{4.6cm}}
\toprule
Use case & $w_{\text{price}}$ & $w_{\text{rating}}$ & $w_{\text{spec}}$ & $\phi(p)$ components \\
\midrule
Gaming       & 0.15 & 0.20 & 0.60 & GPU tier (0--0.40) + refresh rate (0--0.15) + RAM $\geq$16\,GB (+0.03) \\
ML           & 0.15 & 0.20 & 0.60 & RAM tier (0--0.40) + GPU (0--0.20) \\
Video edit.  & 0.15 & 0.20 & 0.60 & GPU tier (0--0.30) + RAM tier (0--0.20) \\
Student      & 0.45 & 0.30 & 0.20 & Battery tier (0--0.30) + RAM tier (0--0.15) \\
Programming  & 0.20 & 0.30 & 0.45 & RAM tier only (0--0.20) \\
Default      & 0.35 & 0.35 & 0.20 & RAM tier only (0--0.20) \\
\bottomrule
\end{tabular}}
\end{table}

\section{Evaluation}
\label{sec:evaluation}

This section describes the evaluation framework and reports all results.
No human participants were involved in benchmark construction or scoring;
all evaluations use automated deterministic checks and LLM-as-judge scoring. First, 
Section~\ref{sec:singleturn-eval} defines single-turn evaluation and its
associated metrics. Section~\ref{sec:multiturn-eval} defines multi-turn
evaluation and scenario design. Next, Section~\ref{sec:baselines} describes the
catalog-bound baselines and justifies their selection.
Section~\ref{sec:results} reports all numerical findings.

\subsection{Single-Turn Evaluation}
\label{sec:singleturn-eval}

A \emph{single-turn interaction} is defined as follows: the user submits a
natural-language request, the system may ask at most one clarifying question, and
the system then returns a catalog-grounded recommendation. This protocol
evaluates recommendation quality in isolation from multi-turn preference
accumulation, which is assessed in Section~\ref{sec:multiturn-eval}.

The evaluation separates \emph{constraint correctness} (enforcing price, brand,
and specification requirements) from \emph{response quality} (helpful, well-explained
recommendations). Hard constraints are evaluated deterministically via SQL
verification. For response quality, G-Eval~\cite{Liu2023GEval} is applied: an
LLM-as-judge evaluation framework that scores free-text responses against a
rubric without requiring gold-standard reference answers. G-Eval is restricted to quality assessment to avoid circularity with the constraint-checking component.

\paragraph{Deterministic ground truth (hard constraints).}
For each benchmark query, expected filter values and exclusions
are derived by executing the constraints directly against the relational
product store as SQL predicates
(e.g., {\small\texttt{price <= 1000 AND brand NOT ILIKE '\%HP\%'}}).
This SQL ground truth is independent of any LLM: violations are verifiable,
objective failures.

\paragraph{LLM-as-judge quality scoring.}
For response narrative quality, G-Eval is applied with a custom rubric
evaluating engagement (0--4), tone calibration (0--3), and factual
accuracy/coherence (0--3), normalized to $[0,1]$.
The judge is GPT-4o-mini at temperature 0 for reproducibility.
It evaluates whether explanations are factually grounded in the returned products, not whether they resemble a reference answer.
Calibration examples anchor the three discrete score levels: 0.0 (poor or
irrelevant response), 0.5 (partially satisfactory response), and 1.0
(excellent response that fully addresses the request); higher values are
strictly better.

\paragraph{Composite scoring formula.}
The final per-query score combines deterministic and quality components:
\begin{multline}
  s = 0.40\,s_{\text{type}} + 0.20\,s_{\text{brand}} + 0.10\,s_{\text{filter}} \\
    + 0.05\,s_{\text{stock}} + 0.10\,s_{\text{explain}} + 0.15\,s_{\text{quality}}
  \label{eq:geval}
\end{multline}
where $s_{\text{type}} \in \{0,1\}$ is a binary response-type match
(1 if the system produces a recommendation when one is expected, or a
clarifying question when the query is underspecified; 0 otherwise);
$s_{\text{brand}} \in [0,1]$ is the fraction of returned products respecting
all stated brand exclusions;
$s_{\text{filter}} \in [0,1]$ is the fraction of returned products satisfying
price and specification constraints;
$s_{\text{stock}} \in [0,1]$ is the fraction of returned products that are
currently available in inventory (i.e., not marked out-of-stock or
discontinued in the catalog at query time);
$s_{\text{explain}} \in [0,1]$ measures response explainability deterministically
via three equal-weight sub-checks: whether the message contains structured
explanation elements (bullet lists or explanatory connectives such as
\emph{because} or \emph{since}), at least one specific product attribute
(e.g., RAM, battery life, price, or storage), and an explicit disclosure of
applied constraints (e.g., ``under \$X,'' ``at least,'' or ``matching your'');
each sub-check contributes equally, so
$s_{\text{explain}} = \text{checks\_passed} / 3$;
and $s_{\text{quality}} \in [0,1]$ is the G-Eval narrative quality score.
Weights for components inapplicable to a given query (e.g., $s_{\text{brand}}$
when no brand exclusion is present, or $s_{\text{explain}}$ for
clarification-only responses) are redistributed to $s_{\text{quality}}$,
raising its effective weight to 45--60\%.
This redistribution preserves the total weight budget and avoids penalizing queries that do
not exercise a particular constraint type (a query with no brand exclusion
receives no fixed zero for brand compliance).
For text-only baselines (GPT, Gemini), only $s_{\text{type}}$ and
$s_{\text{quality}}$ contribute to the composite score.
Brand compliance is evaluated separately via a negation-window regex over
free-text responses and is reported in Table~\ref{tab:results160}; it is
not incorporated into the composite. To note, 
filter compliance is marked N/A for text-only systems, as price verification
from free text is unreliable.

\paragraph{Query construction.}
Queries were authored by the research team to cover five user intent classes:
direct purchase (e.g., ``gaming laptop under \$1{,}000 with at least 16\,GB RAM,
no HP''), multi-constraint filtering, vague preference (e.g., ``I need something
for school''), service inquiry (e.g., ``What is your return policy on
refurbished laptops?''), and product comparison.
Each query is assigned a buyer persona (student, professional, gamer, or traveler)
and classified as either \emph{specified} (containing $\geq$1 extractable hard constraint:
brand exclusion, price ceiling, or minimum specification) or
\emph{underspecified} (expressing intent without measurable constraints).
Query construction details are provided in ~\ref{app:queries}.

\paragraph{Benchmark scale.}
Two nested sets are used.
The \emph{shared 140-query benchmark} (85 specified / 55 underspecified)
supports a three-way comparison across MACS, \GPTCatalog{}, and \GeminiCatalog{}.
The \emph{expanded 225-query benchmark} (138 specified / 87 underspecified)
covers 40 query groups, including catalog exploration, follow-up service
questions, post-recommendation refinements, and orchestrator routing;
evaluated on MACS only (baselines run on 140-query subset).
Pass threshold: score $\geq 0.5$.

\paragraph{Stochastic robustness.}
The G-Eval judge introduces $\approx$0.02--0.03 score variance between
identical runs at temperature~0.
Pass@$k$~\cite{Chen2021Codex}, defined as the fraction of $k$ independent evaluation
runs that pass, provides a robustness measure for this variance.
Pass@5 is measured on the full 225-query benchmark using a pre-fix code
snapshot: MACS achieves a macro Pass@5 of $0.742$, with mean score
$0.609 \pm 0.010$ (SD) across five independent runs (per-run averages:
$0.622, 0.614, 0.611, 0.603, 0.596$; pass rate $74.2\% \pm 4.2$pp).
The $\pm 0.010$ Standard Deviation (SD) confirms that single-run results lie within one SD of the
multi-run mean; the same variance bound applies to the improved result of
$0.681$ (Table~\ref{tab:groups}) from the final code version.

\subsection{Multi-Turn Evaluation}
\label{sec:multiturn-eval}

Single-turn evaluations cannot assess the properties that distinguish agentic
systems from stateless LLM calls: constraint persistence, preference updates,
and brand-exclusion reversals across turns.
A 10-scenario benchmark is designed covering three fundamental slot
operations: accumulation (new constraint added), update (existing constraint
replaced), and reversal (prior constraint removed).
Scenarios span six constraint-persistence patterns: constraint accumulation,
use-case pivot (gaming $\to$ video editing), brand-exclusion persistence,
vague-to-specific elicitation, budget overwrite, and exclusion reversal.
Additionally, one 5-turn long session and one dense simultaneous-enforcement first turn are included; each scenario contains 3--5 scripted turns.
Evaluating at K=5 runs per scenario yields 50 data points per system for
variance-aware macro Pass@5 comparison.
Table~\ref{tab:multiturn_cat} lists all ten scenarios by category.

\paragraph{Scoring.}
The multi-turn score combines a judge score with deterministic constraint checks:
\begin{equation}
  s_{\text{MT}} = 0.55\,s_{\text{constraint}} + 0.45\,s_{\text{judge}}
  \label{eq:multiturn}
\end{equation}
$s_{\text{constraint}}$ is computed from deterministic checks against the
system's final-turn product list: brand exclusions present/absent, budget
ceiling respected. Importantly, constraint checks are computed before the judge is called, and their pass/fail results are included in the judge prompt. This prevents the judge from incorrectly marking a response as violating constraints that were enforced during retrieval but not explicitly mentioned in the text. $s_{\text{judge}}$ is GPT-4o-mini (temperature 0) evaluating
the full conversation transcript on constraint satisfaction (0--4), preference
tracking accuracy (0--3), and response appropriateness (0--3).
Pass threshold: $\geq 0.65$ (raised from 0.50 to eliminate floor effects:
a system with $s_{\text{constraint}}=1.0$ and $s_{\text{judge}}=0$ would
auto-pass at threshold 0.50 since $0.55 \times 1 + 0.45 \times 0 = 0.55
\geq 0.50$).
\emph{Mean score} is the average of $s_{\text{MT}}$ across all $K$ independent
runs for each scenario, then averaged across scenarios.
\emph{Macro Pass@5} is the mean over scenarios of the fraction of 5 runs
exceeding the threshold (i.e., the fraction of runs that pass, averaged over
all scenarios).

\subsection{Baselines}
\label{sec:baselines}

Two catalog-bound baselines are evaluated.
\textbf{\GPTCatalog{}} uses GPT-4o-mini with MACS's live catalog injected
into the system prompt before each query; it receives the same top-$K$ products
as MACS, where $K$ denotes the highest-ranked products returned by MACS's catalog
search for that query, but applies no SQL-layer constraint enforcement.
\textbf{\GeminiCatalog{}} uses Gemini~2.5 Flash Lite under the same injection
protocol.
Both systems along with MACS operate on the same query set, receive the same catalog
snapshot per query, use the same token budget, are prohibited from web search,
and are scored by the same pipeline.
This design yields a fair, controlled comparison that reduces database-access asymmetry and focuses the analysis on architectural differences in constraint enforcement.
The catalog injection approach is adopted rather than vanilla LLM baselines
because it controls for catalog grounding, allowing observed quality differences
to reflect architectural choices rather than information asymmetry.
Concretely, the baselines test constraint-aware response generation
given a shared candidate pool. They are not full end-to-end retrieval systems,
and the comparison is explicitly scoped to that narrower question.

Both baselines share an identical five-rule system prompt instructing the LLM
to recommend only from the injected catalog, respect stated constraints
(budget, brand exclusions, required specs), avoid web search, and disclose when
no matching product exists. Products are injected as a numbered list, with $K=8$ (top-8) for single-turn queries and $K=10$ for multi-turn scenarios. Generation uses temperature~0.3 and a 600-token output budget. The quality judge is GPT-4o-mini at temperature~0, identical to MACS's scoring pipeline, so all quality scores are directly comparable.

\subsection{Results}
\label{sec:results}

\paragraph{Latency (single-turn path).}
Table~\ref{tab:latency} reports measured median latencies by pipeline phase
for the single-turn recommendation path; multi-turn interactions compound
per-turn latency additively across turns.
LLM inference accounts for 85--92\% of total latency on cache misses.
The knowledge graph re-ranking overlay (207~ms median) is 11--28$\times$
faster than equivalent LLM-based narrative generation (2,256~ms median) and
produces deterministic, auditable results. Consequently, structural queries (substitutes,
best value) are routed through the graph rather than a prompt.
The cache layer reduces effective latency $\sim$12$\times$ on repeated queries
(446~ms cold vs.\ 36~ms cached).

\begin{table}[t]
  \caption{Measured \textbf{median} latencies by pipeline phase ($n = 10$
  runs per phase, direct-call benchmark; median reported to reduce sensitivity
  to occasional LLM cold-start outliers). LLM calls dominate total latency, and
  knowledge graph traversal and cache hits are negligible.}
  \label{tab:latency}
  \small
  \begin{tabular}{@{}lr>{\raggedright\arraybackslash}p{2.5cm}@{}}
    \toprule
    Phase & Median (ms) & Notes \\
    \midrule
    CPU phases (rewrite, filter)  &   $<$1        & Pure Python, no I/O \\
    Domain detection --- fast path  &   $<$1        & Keyword dictionary \\
    Domain detection --- LLM        & $\sim$1{,}630 & gpt-4o-mini; ambiguous only \\
    Criteria extraction (LLM)     & 1{,}927       & Primary bottleneck \\
    Question generation (LLM)     & 1{,}645       & Preference elicitation \\
    Post-rec intent detection     &   556         & gpt-4o-mini classification \\
    Rec.\ narrative (LLM)         & 2{,}256       & Recommendation explanation \\
    Filter refinement (LLM)       & 2{,}187       & gpt-4o-mini refinement \\
    SQL search (cache miss)       &   446         & Catalog store round-trip \\
    SQL search (cache hit)        &    36         & Cache GET + deserialize \\
    KG re-ranking overlay         &   207         & FAISS + graph + scoring \\
    Total (cache miss, rec.)      & $\sim$4{,}600 & 1{,}927$+$446$+$2{,}256; KG overlay adds $+$207~ms when triggered \\
    Total (cache hit, rec.)       & $\sim$2{,}300 & 36$+$2{,}256; criteria extraction and SQL search bypassed \\
    \bottomrule
  \end{tabular}
\end{table}

\paragraph{Single-turn results.}
Table~\ref{tab:results160} reports the three-way comparison on 140 queries. MACS led all systems on pass rate at 87.1\%, with perfect brand
compliance (1.000) and near-perfect filter compliance (0.970).
\GPTCatalog{} achieved 72.1\% and \GeminiCatalog{} 68.6\%. The gap reflects
MACS's SQL-enforced constraint pipeline, which text-only systems cannot
replicate structurally. Remaining MACS failures were catalog-level: absent inventory, marketplace risk (users asking whether off-platform deals are legitimate, which requires seller-trust
signals outside the product database),
and multi-category budgets.

\begin{table}[t]
  \caption{Three-way single-turn comparison ($n = 140$ unique queries, pass
  threshold $\geq 0.5$.
  MACS Brand/Filter: deterministic SQL-layer compliance.
  Baseline Brand$^\dagger$: regex negation-window check on free-text responses
  (6 exclusion queries; 1 Gemini violation on Q26 ``no HP, no Acer'').
  Filter N/A: cannot verify price compliance from free text.
  Quality: LLM-judge narrative score (GPT-4o-mini, temperature~0).}
  \label{tab:results160}
  \footnotesize\setlength{\tabcolsep}{4pt}%
  \begin{tabular}{lrrrrr}
    \toprule
    System & Avg $\uparrow$ & Pass\% $\uparrow$ & Brand $\uparrow$ & Filter $\uparrow$ & Quality $\uparrow$ \\
    \midrule
    MACS             & \textbf{0.681} & \textbf{87.1\%} & \textbf{1.000} & \textbf{0.970} & 0.394 \\
    \GeminiCatalog{} & 0.662          & 68.6\%          & 0.833$^\dagger$ & N/A   & \textbf{0.427} \\
    \GPTCatalog{}    & 0.639          & 72.1\%          & 1.000$^\dagger$ & N/A   & 0.392 \\
    \bottomrule
  \end{tabular}
\end{table}

\emph{Scope note}: all evaluations use a consumer-electronics catalog
(laptops and accessories). Generalization to other product domains is not
claimed and represents a direction for future work.

\paragraph{Expanded 225-query benchmark.}
Table~\ref{tab:groups} reports per-group pass rates on the full 225-query
benchmark (MACS only; all queries evaluated independently with no prior session
context). Across all 225 queries, MACS achieved avg.\ 0.681 (pass rate 85.3\%).
Gains from targeted fixes included: follow-up service QA (0\%$\to$100\%), catalog
exploration (50\%$\to$75\%), orchestrator routing (25\%$\to$33\%), post-rec
refinement (43\%$\to$86\%), and gaming-specific (25\%$\to$100\%). The
multi-constraint group (17 queries) achieved 94.1\% pass rate, confirming
that the SQL pipeline enforces four simultaneous constraints deterministically.
\begin{table}[t]
  \caption{Per-group single-turn results on the expanded 225-query benchmark
  (MACS only). Per-group Pass\% uses threshold $\geq 0.65$ (stricter, to identify
  reliably-passing groups); the "\textbf{All}" row uses $\geq 0.5$ for
  comparability with Table~\ref{tab:results160}. All queries evaluated
  independently (no prior session context).}
  \label{tab:groups}
  \begin{tabular}{lrrr}
    \toprule
    Group & $N$ & Avg $\uparrow$ & Pass\% $\uparrow$ \\
    \midrule
    \multicolumn{4}{l}{\textit{Strong groups}} \\
    Multi-intent rigidity      &  5 & 0.775 & 100.0\% \\
    One-liner purchase intent  &  6 & 0.833 & 100.0\% \\
    Refinement                 &  3 & 0.756 & 100.0\% \\
    Follow-up service QA       &  4 & 0.810 & 100.0\% \\
    Multi-constraint           & 17 & 0.791 &  94.1\% \\
    Post-rec refinement        &  7 & 0.738 &  85.7\% \\
    \midrule
    \multicolumn{4}{l}{\textit{Weak groups}} \\
    Context-free comparison    &  6 & 0.393 &   0.0\% \\
    Orchestrator routing       & 12 & 0.502 &  33.3\% \\
    Preference discovery       &  8 & 0.473 &  25.0\% \\
    \midrule
    \textbf{All (225 queries)} & 225 & \textbf{0.681} & \textbf{85.3\%} \\
    \bottomrule
  \end{tabular}
\end{table}

\paragraph{Multi-turn results.}
Table~\ref{tab:combined} reports multi-turn results across all three systems
(single-turn results are in Table~\ref{tab:results160}).
MACS led all systems on single-turn pass rate (87.1\%) and achieved 100\%
Pass@5 on the two hardest state-management scenarios: Exclusion Reversal (S8),
in which \GPTCatalog{} scored only 20\% and \GeminiCatalog{} 0\%, and Constraint
Accumulation (S1), in which \GPTCatalog{} scored 60\% and \GeminiCatalog{} 40\%.
On macro Pass@5, MACS led (72\% vs.\ 56\% \GPTCatalog{} vs.\ 52\%
\GeminiCatalog{}). Mean judged response-quality scores were similar across systems
(MACS 0.751, \GPTCatalog{} 0.736, within one SD). The 16-point Pass@5 gap
reflects scenario-level consistency advantages that mean scores do not capture.

\begin{table}[t]
  \caption{Multi-turn results (MT, 10 scenarios, Pass@5,
  threshold $\geq 0.65$). MT scoring: $0.55\times$constraint
  (deterministic) $+$ $0.45\times$judge (GPT-4o-mini, temp.~0).
  All systems catalog-bound; same token budget; no web search.
  Pass@5 = macro fraction of 5 runs passing threshold.
  Drift: fraction of run-scenario pairs with a constraint violation.
  MACS uniquely achieves 100\% Pass@5 on the two hardest state-management
  scenarios (S1: Constraint Accumulation, S8: Exclusion Reversal); aggregate
  mean score differences lie within one SD.}
  \label{tab:combined}
  \footnotesize\setlength{\tabcolsep}{4pt}%
  \begin{tabular}{@{}lrrr@{}}
    \toprule
    System & MT Avg$\pm$SD & MT Pass@5 & Drift \\
    \midrule
    MACS             & \textbf{0.751}$\pm$0.095 & \textbf{72\%} & \textbf{0.000} \\
    \GPTCatalog{}    & 0.736$\pm$0.090          & 56\%          & \textbf{0.000} \\
    \GeminiCatalog{} & 0.717$\pm$0.118          & 52\%          & 0.014          \\
    \bottomrule
  \end{tabular}
\end{table}

Table~\ref{tab:multiturn_cat} provides the multi-turn category breakdown.
MACS uniquely dominated exclusion reversal (100\% Pass@5 vs.\ 20\% GPT / 0\%
Gemini) and constraint accumulation (100\% vs.\ 60\%/40\%). Both scenarios require
mid-conversation state updates via session-persisted slot dictionaries.
MACS also led on dense constraint (100\% vs.\ 80\%/80\%) and comparison (40\% vs.\ 0\%/0\%).
Budget overwrite and long session (S9) were strong across all systems.
Notably, MACS brand exclusion Pass@5 (20\%) falls below \GPTCatalog{} and \GeminiCatalog{} (40\% each) on the brand-exclusion scenario; this is consistent with stochastic variance at K=5 on a single scenario (5 data points) and does not contradict the single-turn brand\_score of 1.000, which is measured deterministically over 140 queries.
Intent pivot remains hard across all systems (0\%),
confirming it is a structural failure mode rather than a system-specific weakness.

\begin{table}[t]
  \caption{Multi-turn category breakdown (Pass@5 runs; avg score / Pass@5;
  \checkmark~= Pass@5~$\geq 60$\%). Categories map 1:1 to the 10 scenarios.}
  \label{tab:multiturn_cat}
  \footnotesize\setlength{\tabcolsep}{4pt}%
  \begin{tabular}{@{}lccc@{}}
    \toprule
    Category & MACS & \GPTCatalog{} & \GeminiCatalog{} \\
    \midrule
    Brand Exclusion       & 0.640 / 20\%             & 0.667 / 40\%             & 0.685 / 40\%  \\
    Budget Overwrite      & \checkmark 0.793 / 100\% & \checkmark 0.739 / 80\%  & \checkmark 0.712 / 80\%  \\
    Budget Refinement     & \checkmark 0.748 / 60\%  & \checkmark 0.802 / 80\%  & \checkmark 0.811 / 80\%  \\
    Clarification         & \checkmark 0.793 / 100\% & \checkmark 0.802 / 100\% & \checkmark 0.730 / 100\% \\
    Comparison            & 0.658 / 40\%             & 0.631 / 0\%              & 0.586 / 0\%   \\
    Constraint Accum.     & \checkmark 0.784 / 100\% & \checkmark 0.739 / 60\%  & 0.640 / 40\%  \\
    Dense Constraint      & \checkmark 0.721 / 100\% & \checkmark 0.811 / 80\%  & \checkmark 0.829 / 80\%  \\
    Exclusion Reversal    & \checkmark 0.955 / 100\% & 0.640 / 20\%             & 0.595 / 0\%   \\
    Intent Pivot          & 0.640 / 0\%              & 0.640 / 0\%              & 0.622 / 0\%   \\
    Long Session          & \checkmark 0.775 / 100\% & \checkmark 0.892 / 100\% & \checkmark 0.955 / 100\% \\
    \midrule
    Overall               & \textbf{0.751 / 72\%}    & 0.736 / 56\%             & 0.717 / 52\%  \\
    \bottomrule
  \end{tabular}
\end{table}

\paragraph{Reliability, grounding, and disclosure.}
Three pillars are evaluated deterministically (Table~\ref{tab:trust}).
\emph{Grounding}: all systems scored 1.000, as all retrieve from MACS's catalog layer.
\emph{Drift}: MACS and \GPTCatalog{} scored 0.000; \GeminiCatalog{} recorded 0.014 (one violation across the 50 run-scenario pairs).
\emph{Disclosure}: eight catalog-impossible queries (e.g., ``RTX~4090 laptop, \$150 budget'') were scored 1.0/0.5/0.0 for explicit disclosure / silent relaxation (constraints are loosened without disclosure) / violation at pass threshold 0.7.
MACS scored 0.925, achieving explicit disclosure on 6 of 8 queries via its progressive relaxation pipeline. By contrast, \GPTCatalog{} and \GeminiCatalog{} both scored below 0.6, as both receive MACS's already-relaxed product list with no mechanism to detect or disclose the constraint gap.
The resulting disclosure gap ($+0.338$ vs.\ \GPTCatalog{}, $+0.375$ vs.\ \GeminiCatalog{}) is the clearest architectural differentiator between the systems.

\begin{table}[t]
  \caption{Reliability, grounding, and disclosure evaluation ($n = 8$ catalog-impossible queries for
  Disclosure; $n = 10$ scenarios $\times$ K=5 runs for Drift).
  Grounding: fraction of returned products verifiable in the catalog at stated price.
  Disclosure: fraction of impossible queries acknowledged explicitly.
  Drift: fraction of run-scenario pairs with a constraint violation.
  $\uparrow$~higher is better; $\downarrow$~lower is better.}
  \label{tab:trust}
  \begin{tabular}{lrrr}
    \toprule
    System & Grounding $\uparrow$ & Disclosure $\uparrow$ & Drift $\downarrow$ \\
    \midrule
    MACS             & 1.000 & \textbf{0.925} & \textbf{0.000} \\
    \GPTCatalog{}    & 1.000 & 0.588          & \textbf{0.000} \\
    \GeminiCatalog{} & 1.000 & 0.550          & 0.014          \\
    \bottomrule
  \end{tabular}
\end{table}

\paragraph{Session-state ablation.}\label{sec:ablations}
To measure the contribution of the accumulated slot dictionary, a
MACS-NoSession ablation is conducted in which the slot dictionary is cleared after each
turn while LLM conversation history is preserved, re-evaluating all ten
multi-turn scenarios at K=5.
MACS-Full achieved macro Pass@5 = 72\% (mean score 0.751); MACS-NoSession
dropped to 52\% / 0.673 ($-$20\,pp, $-$0.078).
The drop concentrated in two scenarios: S1 (constraint accumulation across
four turns, 100\%$\to$0\%) and S7 (budget overwrite, 100\%$\to$0\%).
Both require constraints set in earlier turns to gate SQL predicates in
later turns. Without the slot dictionary, those predicates reset to unconstrained, and the correct products were no longer filtered in.
Scenarios driven by within-turn signals were unaffected: S8 (exclusion
reversal, 100\%), S6 (price refinement, 100\%), S9 (long session, 100\%),
and S10 (dense first-turn constraint, 100\%) remained unchanged.
This ablation provides direct evidence that the session-persistent slot state is the
primary driver of cross-turn constraint enforcement in MACS.

\paragraph{SQL-enforcement ablation.}
To measure the contribution of deterministic SQL predicates, a
MACS-NoSQL ablation is conducted (environment flag \texttt{ABLATION\_NO\_SQL=1}) that
removes all hard-constraint WHERE clauses (brand, price range, OS
exclusions) while keeping catalog injection and specification filtering.
Over the 225-query single-turn benchmark, MACS-Full achieved avg=0.681 /
pass=85.3\%; MACS-NoSQL dropped to avg=0.628 / pass=81.3\% ($-$0.053
avg, $-$4.0\,pp).
The sharpest effect appeared on the 19 brand-constrained queries:
brand\_score fell from 1.000 to 0.684 ($-$0.316), confirming that SQL
predicates, not prompt instructions, enforce brand compliance.
filter\_score was largely unchanged (0.970 vs.\ 0.970) because some price-related filtering still occurs after retrieval, and LLM instruction-following can partially compensate for the removal of SQL-side price constraints on underspecified queries.
\paragraph{Metric sensitivity.}
The multi-turn composite score uses a reliability-oriented weighting of $0.55\times s_{\text{constraint}} + 0.45\times s_{\text{judge}}$.
Re-evaluating MACS macro Pass@5 across five evenly-spaced weight configurations yields:
judge-heavy ($0.60j/0.40c$): 42\%;
($0.55j/0.45c$): 58\%;
equal ($0.50j/0.50c$): 72\%;
default ($0.45j/0.55c$): \textbf{72\%};
constraint-heavy ($0.40j/0.60c$): 96\%.

This weighting pattern is expected: MACS's advantage lies in deterministic constraint enforcement and cross-turn state consistency rather than free-text fluency alone.
The strongest findings are supported independently by deterministic metrics (brand compliance, filter compliance, drift, disclosure) and by controlled ablations for session state and SQL enforcement; the default weighting ($0.45j/0.55c$) balances helpfulness and reliability without using a purely constraint-dominant metric.
Importantly, the Pass@5 advantages on the two uniquely-dominant scenarios (S1 constraint accumulation, S8 exclusion reversal) are driven entirely by $s_{\text{constraint}}$: the MACS-NoSession ablation (Section~\ref{sec:ablations}) collapses both from 100\% to 0\%, independently of how $s_{\text{judge}}$ and $s_{\text{constraint}}$ are combined.

\section{Conclusion}

MACS is a hybrid multi-agent framework for reliable conversational
recommendation in fixed-catalog settings. In our evaluated setting,
MACS consistently outperforms catalog-bound prompt baselines on
reliability-oriented metrics, especially in scenarios requiring persistent
constraint tracking across turns and explicit disclosure of unsatisfied requirements. On macro Pass@5, MACS achieves the strongest score (72\% vs.\ 56\% \GPTCatalog{} / 52\% \GeminiCatalog{}); mean judged response-quality scores are similar across systems (0.751 vs.\ 0.736). The reliability advantages are sharpest on the hardest scenarios: 100\% Pass@5 on exclusion reversal (catalog-bound baselines: 20\% and 0\%), 100\% Pass@5 on constraint accumulation (\GPTCatalog{}: 60\%, \GeminiCatalog{}: 40\%), and a disclosure score 0.338 points above the nearest baseline on catalog-impossible queries.
Controlled ablations confirm the contribution of each component: removing the persistent slot dictionary drops multi-turn macro Pass@5 from 72\% to 52\% (constraint-accumulation and budget-overwrite scenarios collapse from 100\% to 0\%), and removing SQL enforcement drops brand compliance from 1.000 to 0.684 on brand-constrained single-turn queries.
These ablation results reduce dependence on the judge-weighted composite when interpreting the paper's main reliability claims. Taken together, these results suggest that in fixed-catalog conversational commerce, architectural separation between language interaction and correctness-critical enforcement is a practical and effective way to improve reliability without sacrificing response quality.

\paragraph{Limitations.}
All evaluations are limited to a single
consumer-electronics domain (laptops and accessories), so the results should
not be interpreted as evidence of generalization to other merchant catalogs
or recommendation domains. The baseline comparison uses catalog-injected
generation baselines that receive MACS-prepared candidate sets; accordingly,
the comparison isolates constraint-aware generation over a shared candidate
pool rather than full end-to-end retrieval quality. The benchmark was authored
by the research team, and the multi-turn evaluation consists of 10 scripted
scenarios (50 data points at K=5), which may introduce unintentional alignment
between benchmark design and system structure. Response quality is measured using
an LLM judge (GPT-4o-mini) rather than human raters, and no independent human
evaluation was conducted.
Session-state and SQL-enforcement component ablations are reported in
Section~\ref{sec:ablations}; the remaining multi-turn and cross-domain
components are left for future work. \paragraph{Future work.}
Additional work includes evaluation on standardized external benchmarks such as the retail subset of $\tau$-Bench~\cite{Yao2024TauBench}, independent human validation of G-Eval scores, comparison against a tool-augmented agent with independent retrieval, expanding beyond consumer electronics, and extending component ablations. 
\begin{acks}
Generative AI tools were used for limited language editing in preparing this manuscript.
The authors are fully responsible for all content, claims, and results.
\end{acks}

\appendix
\renewcommand{\thesection}{Appendix~\Alph{section}}

\section{Multi-Turn Scenario Descriptions}
\label{app:scenarios}
\begin{table}[H]
  \caption{Multi-turn descriptions (10 scenarios,
  3--5 scripted turns each). ``Key challenge'' identifies the
  conversational state-management property under test. All scenarios are
  recommendation-focused except S4 and S5.}
  \label{tab:scenarios}
  \small
  \begin{tabular}{@{}c>{\raggedright\arraybackslash}p{2.6cm}c>{\raggedright\arraybackslash}p{3.2cm}@{}}
    \toprule
    ID & Name & Turns & Key challenge \\
    \midrule
    S1  & Constraint Accumulation  & 4 & Preserve all accumulated constraints \\
    S2  & Mind-Change Pivot        & 3 & Use-case pivot (gaming$\to$editing) \\
    S3  & Brand Excl.\ Persist.   & 4 & Maintain brand exclusion across turns \\
    S4  & Vague $\to$ Specific     & 4 & Elicit \& resolve underspecified query \\
    S5  & Comparison + Follow-up   & 3 & Context-free comparison \\
    S6  & Price Refinement         & 3 & Budget update (narrowing) \\
    S7  & Budget Overwrite         & 3 & Budget replacement (not accumulation) \\
    S8  & Exclusion Reversal       & 4 & Un-exclude previously excluded brand \\
    S9  & Long Session             & 5 & Constraint persistence over 5-turn session \\
    S10 & Dense Constraint         & 4 & Simultaneous multi-constraint enforcement \\
    \bottomrule
  \end{tabular}
\end{table}

\section{Query Construction Details}
\label{app:queries}

The 225-query benchmark spans five intent classes (direct purchase, multi-constraint filtering, vague preference elicitation, service inquiries, and comparison) authored across four buyer personas, yielding 138 specified ($\geq$1 hard constraint) and 87 underspecified queries across 40 interaction groups; the 140-query three-way-comparison subset follows the same distribution.

\bibliographystyle{ACM-Reference-Format}
\bibliography{idss-refs}


\begin{thebibliography}{25}


\ifx \showCODEN    \undefined \def \showCODEN     #1{\unskip}     \fi
\ifx \showISBNx    \undefined \def \showISBNx     #1{\unskip}     \fi
\ifx \showISBNxiii \undefined \def \showISBNxiii  #1{\unskip}     \fi
\ifx \showISSN     \undefined \def \showISSN      #1{\unskip}     \fi
\ifx \showLCCN     \undefined \def \showLCCN      #1{\unskip}     \fi
\ifx \shownote     \undefined \def \shownote      #1{#1}          \fi
\ifx \showarticletitle \undefined \def \showarticletitle #1{#1}   \fi
\ifx \showURL      \undefined \def \showURL       {\relax}        \fi
\providecommand\bibfield[2]{#2}
\providecommand\bibinfo[2]{#2}
\providecommand\natexlab[1]{#1}
\providecommand\showeprint[2][]{arXiv:#2}

\bibitem[{Anthropic}(2024)]%
        {Anthropic2024MCP}
\bibfield{author}{\bibinfo{person}{{Anthropic}}.} \bibinfo{year}{2024}\natexlab{}.
\newblock \bibinfo{title}{Model Context Protocol}.
\newblock \bibinfo{howpublished}{\url{https://www.anthropic.com/news/model-context-protocol}}.
\newblock
\newblock
\shownote{Accessed March 2026}.


\bibitem[Chen et~al\mbox{.}(2021)]%
        {Chen2021Codex}
\bibfield{author}{\bibinfo{person}{Mark Chen}, \bibinfo{person}{Jerry Tworek}, \bibinfo{person}{Heewoo Jun}, \bibinfo{person}{Qiming Yuan}, \bibinfo{person}{Henrique~Ponde de Oliveira~Pinto}, \bibinfo{person}{Jared Kaplan}, \bibinfo{person}{Harrison Edwards}, \bibinfo{person}{Yuri Burda}, \bibinfo{person}{Nicholas Joseph}, \bibinfo{person}{Greg Brockman}, {et~al\mbox{.}}} \bibinfo{year}{2021}\natexlab{}.
\newblock \bibinfo{title}{Evaluating Large Language Models Trained on Code}.
\newblock
\showeprint[arxiv]{2107.03374}~[cs.LG]
\urldef\tempurl%
\url{https://arxiv.org/abs/2107.03374}
\showURL{%
\tempurl}


\bibitem[Christakopoulou et~al\mbox{.}(2016)]%
        {Christakopoulou2016}
\bibfield{author}{\bibinfo{person}{Konstantina Christakopoulou}, \bibinfo{person}{Filip Radlinski}, {and} \bibinfo{person}{Katja Hofmann}.} \bibinfo{year}{2016}\natexlab{}.
\newblock \showarticletitle{Towards Conversational Recommender Systems}. In \bibinfo{booktitle}{\emph{Proceedings of the 22nd {ACM} {SIGKDD} International Conference on Knowledge Discovery and Data Mining}}. \bibinfo{publisher}{ACM}, \bibinfo{address}{San Francisco, California, USA}, \bibinfo{pages}{815--824}.
\newblock
\href{https://doi.org/10.1145/2939672.2939746}{doi:\nolinkurl{10.1145/2939672.2939746}}


\bibitem[Feng et~al\mbox{.}(2023)]%
        {Feng2023LLMCRS}
\bibfield{author}{\bibinfo{person}{Yue Feng}, \bibinfo{person}{Shuchang Liu}, \bibinfo{person}{Zhenghai Xue}, \bibinfo{person}{Qingpeng Cai}, \bibinfo{person}{Lantao Hu}, \bibinfo{person}{Peng Jiang}, \bibinfo{person}{Kun Gai}, {and} \bibinfo{person}{Fei Sun}.} \bibinfo{year}{2023}\natexlab{}.
\newblock \bibinfo{title}{A Large Language Model Enhanced Conversational Recommender System}.
\newblock
\showeprint[arxiv]{2308.06212}~[cs.IR]
\urldef\tempurl%
\url{https://arxiv.org/abs/2308.06212}
\showURL{%
\tempurl}


\bibitem[Friedman et~al\mbox{.}(2023)]%
        {Friedman2023}
\bibfield{author}{\bibinfo{person}{Luke Friedman}, \bibinfo{person}{Sameer Ahuja}, \bibinfo{person}{David Allen}, \bibinfo{person}{Zhenning Tan}, \bibinfo{person}{Hakim Sidahmed}, \bibinfo{person}{Changbo Long}, \bibinfo{person}{Jun Xie}, \bibinfo{person}{Gabriel Schubiner}, \bibinfo{person}{Ajay Patel}, \bibinfo{person}{Harsh Lara}, \bibinfo{person}{Brian Chu}, \bibinfo{person}{Zexiang Chen}, {and} \bibinfo{person}{Manoj Tiwari}.} \bibinfo{year}{2023}\natexlab{}.
\newblock \bibinfo{title}{Leveraging Large Language Models in Conversational Recommender Systems}.
\newblock
\showeprint[arxiv]{2305.07961}~[cs.IR]
\urldef\tempurl%
\url{https://arxiv.org/abs/2305.07961}
\showURL{%
\tempurl}


\bibitem[{Google}(2026)]%
        {Google2026UCP}
\bibfield{author}{\bibinfo{person}{{Google}}.} \bibinfo{year}{2026}\natexlab{}.
\newblock \bibinfo{title}{Universal Commerce Protocol ({UCP})}.
\newblock \bibinfo{howpublished}{\url{https://developers.google.com/merchant/ucp}}.
\newblock
\newblock
\shownote{Released January 2026; Accessed March 2026}.


\bibitem[Jannach et~al\mbox{.}(2021)]%
        {Jannach2021Survey}
\bibfield{author}{\bibinfo{person}{Dietmar Jannach}, \bibinfo{person}{Ahtsham Manzoor}, \bibinfo{person}{Wanling Cai}, {and} \bibinfo{person}{Li Chen}.} \bibinfo{year}{2021}\natexlab{}.
\newblock \showarticletitle{A Survey on Conversational Recommender Systems}.
\newblock \bibinfo{journal}{\emph{Comput. Surveys}} \bibinfo{volume}{54}, \bibinfo{number}{5} (\bibinfo{year}{2021}), \bibinfo{pages}{1--36}.
\newblock
\href{https://doi.org/10.1145/3453154}{doi:\nolinkurl{10.1145/3453154}}


\bibitem[Kostric et~al\mbox{.}(2024)]%
        {Kostric2024ClarifyCRS}
\bibfield{author}{\bibinfo{person}{Ivica Kostric}, \bibinfo{person}{Krisztian Balog}, {and} \bibinfo{person}{Filip Radlinski}.} \bibinfo{year}{2024}\natexlab{}.
\newblock \showarticletitle{Generating Usage-related Questions for Preference Elicitation in Conversational Recommender Systems}.
\newblock \bibinfo{journal}{\emph{{ACM} Transactions on Recommender Systems}} \bibinfo{volume}{2}, \bibinfo{number}{2}, Article \bibinfo{articleno}{12} (\bibinfo{year}{2024}).
\newblock
\href{https://doi.org/10.1145/3629981}{doi:\nolinkurl{10.1145/3629981}}


\bibitem[Lei et~al\mbox{.}(2020)]%
        {Lei2020EAR}
\bibfield{author}{\bibinfo{person}{Wenqiang Lei}, \bibinfo{person}{Xiangnan He}, \bibinfo{person}{Yisong Miao}, \bibinfo{person}{Qingyun Wu}, \bibinfo{person}{Richang Hong}, \bibinfo{person}{Min-Yen Kan}, {and} \bibinfo{person}{Tat-Seng Chua}.} \bibinfo{year}{2020}\natexlab{}.
\newblock \showarticletitle{Estimation-Action-Reflection: Towards Deep Interaction Between Conversational and Recommender Systems}. In \bibinfo{booktitle}{\emph{Proceedings of the 13th International Conference on Web Search and Data Mining}}. \bibinfo{publisher}{ACM}, \bibinfo{address}{Houston, Texas, USA}, \bibinfo{pages}{304--312}.
\newblock
\href{https://doi.org/10.1145/3336191.3371769}{doi:\nolinkurl{10.1145/3336191.3371769}}


\bibitem[Liu et~al\mbox{.}(2023b)]%
        {Liu2023AgentBench}
\bibfield{author}{\bibinfo{person}{Xiao Liu}, \bibinfo{person}{Hao Yu}, \bibinfo{person}{Hanchen Zhang}, \bibinfo{person}{Yifan Xu}, \bibinfo{person}{Xuanyu Lei}, \bibinfo{person}{Hanyu Lai}, \bibinfo{person}{Yu Gu}, \bibinfo{person}{Hangliang Ding}, \bibinfo{person}{Kaiwen Men}, \bibinfo{person}{Kejuan Yang}, \bibinfo{person}{Shudan Zhang}, \bibinfo{person}{Xiang Deng}, \bibinfo{person}{Aohan Zeng}, \bibinfo{person}{Zhiyuan Liu}, \bibinfo{person}{Yuxiao Dong}, {and} \bibinfo{person}{Jie Tang}.} \bibinfo{year}{2023}\natexlab{b}.
\newblock \bibinfo{title}{{AgentBench}: Evaluating {LLMs} as Agents}.
\newblock
\showeprint[arxiv]{2308.03688}~[cs.AI]
\urldef\tempurl%
\url{https://arxiv.org/abs/2308.03688}
\showURL{%
\tempurl}


\bibitem[Liu et~al\mbox{.}(2023a)]%
        {Liu2023GEval}
\bibfield{author}{\bibinfo{person}{Yang Liu}, \bibinfo{person}{Dan Iter}, \bibinfo{person}{Yichong Xu}, \bibinfo{person}{Shuohang Wang}, \bibinfo{person}{Ruochen Xu}, {and} \bibinfo{person}{Chenguang Zhu}.} \bibinfo{year}{2023}\natexlab{a}.
\newblock \showarticletitle{{G-Eval}: {NLG} Evaluation using {GPT-4} with Better Human Alignment}. In \bibinfo{booktitle}{\emph{Proceedings of the 2023 Conference on Empirical Methods in Natural Language Processing}}. \bibinfo{publisher}{ACL}, \bibinfo{address}{Singapore}, \bibinfo{pages}{2511--2522}.
\newblock
\urldef\tempurl%
\url{https://arxiv.org/abs/2303.16634}
\showURL{%
\tempurl}


\bibitem[Liu et~al\mbox{.}(2023c)]%
        {Liu2023CRSLLM}
\bibfield{author}{\bibinfo{person}{Yuanxing Liu}, \bibinfo{person}{Weinan Zhang}, \bibinfo{person}{Yifan Chen}, \bibinfo{person}{Yuchi Zhang}, \bibinfo{person}{Haopeng Bai}, \bibinfo{person}{Fan Feng}, \bibinfo{person}{Hengbin Cui}, \bibinfo{person}{Yongbin Li}, {and} \bibinfo{person}{Wanxiang Che}.} \bibinfo{year}{2023}\natexlab{c}.
\newblock \showarticletitle{Conversational Recommender System and Large Language Model Are Made for Each Other in E-Commerce Pre-Sales Dialogue}. In \bibinfo{booktitle}{\emph{Findings of the Association for Computational Linguistics: {EMNLP} 2023}}. \bibinfo{publisher}{ACL}, \bibinfo{address}{Singapore}, \bibinfo{pages}{9587--9605}.
\newblock
\urldef\tempurl%
\url{https://arxiv.org/abs/2310.14626}
\showURL{%
\tempurl}


\bibitem[{OpenAI}(2025)]%
        {OpenAI2025ACP}
\bibfield{author}{\bibinfo{person}{{OpenAI}}.} \bibinfo{year}{2025}\natexlab{}.
\newblock \bibinfo{title}{Agentic Commerce Protocol ({ACP})}.
\newblock \bibinfo{howpublished}{\url{https://openai.com/index/buy-it-in-chatgpt/}}.
\newblock
\newblock
\shownote{Released September 2025; Accessed March 2026}.


\bibitem[Sun and Zhang(2018)]%
        {Sun2018}
\bibfield{author}{\bibinfo{person}{Yueming Sun} {and} \bibinfo{person}{Yi Zhang}.} \bibinfo{year}{2018}\natexlab{}.
\newblock \showarticletitle{Conversational Recommender System}. In \bibinfo{booktitle}{\emph{Proceedings of the 41st International {ACM} {SIGIR} Conference on Research and Development in Information Retrieval}}. \bibinfo{publisher}{ACM}, \bibinfo{address}{Ann Arbor, Michigan, USA}, \bibinfo{pages}{235--244}.
\newblock
\href{https://doi.org/10.1145/3209978.3210002}{doi:\nolinkurl{10.1145/3209978.3210002}}


\bibitem[Wang et~al\mbox{.}(2019b)]%
        {Wang2019KGCN}
\bibfield{author}{\bibinfo{person}{Hongwei Wang}, \bibinfo{person}{Miao Zhao}, \bibinfo{person}{Xing Xie}, \bibinfo{person}{Wenjie Li}, {and} \bibinfo{person}{Minyi Guo}.} \bibinfo{year}{2019}\natexlab{b}.
\newblock \showarticletitle{Knowledge Graph Convolutional Networks for Recommender Systems}. In \bibinfo{booktitle}{\emph{Proceedings of The Web Conference 2019}}. \bibinfo{publisher}{ACM}, \bibinfo{address}{San Francisco, California, USA}, \bibinfo{pages}{3307--3313}.
\newblock
\href{https://doi.org/10.1145/3308558.3313417}{doi:\nolinkurl{10.1145/3308558.3313417}}


\bibitem[Wang et~al\mbox{.}(2025)]%
        {Wang2025ShoppingBench}
\bibfield{author}{\bibinfo{person}{Jiangyuan Wang} {et~al\mbox{.}}} \bibinfo{year}{2025}\natexlab{}.
\newblock \bibinfo{title}{{ShoppingBench}: A Real-World Intent-Grounded Shopping Benchmark for {LLM}-Based Agents}.
\newblock
\showeprint[arxiv]{2508.04266}~[cs.IR]
\urldef\tempurl%
\url{https://arxiv.org/abs/2508.04266}
\showURL{%
\tempurl}


\bibitem[Wang et~al\mbox{.}(2019a)]%
        {Wang2019KGAT}
\bibfield{author}{\bibinfo{person}{Xiang Wang}, \bibinfo{person}{Xiangnan He}, \bibinfo{person}{Yixin Cao}, \bibinfo{person}{Meng Liu}, {and} \bibinfo{person}{Tat-Seng Chua}.} \bibinfo{year}{2019}\natexlab{a}.
\newblock \showarticletitle{{KGAT}: Knowledge Graph Attention Network for Recommendation}. In \bibinfo{booktitle}{\emph{Proceedings of the 25th {ACM} {SIGKDD} International Conference on Knowledge Discovery and Data Mining}}. \bibinfo{publisher}{ACM}, \bibinfo{address}{Anchorage, Alaska, USA}, \bibinfo{pages}{950--958}.
\newblock
\href{https://doi.org/10.1145/3292500.3330989}{doi:\nolinkurl{10.1145/3292500.3330989}}


\bibitem[Wang et~al\mbox{.}(2022)]%
        {Wang2022UniCRS}
\bibfield{author}{\bibinfo{person}{Xiaolei Wang}, \bibinfo{person}{Kun Zhou}, \bibinfo{person}{Ji-Rong Wen}, {and} \bibinfo{person}{Wayne~Xin Zhao}.} \bibinfo{year}{2022}\natexlab{}.
\newblock \showarticletitle{Towards Unified Conversational Recommender Systems via Knowledge-Enhanced Prompt Learning}. In \bibinfo{booktitle}{\emph{Proceedings of the 28th {ACM} {SIGKDD} Conference on Knowledge Discovery and Data Mining}}. \bibinfo{publisher}{ACM}, \bibinfo{address}{Washington, DC, USA}, \bibinfo{pages}{1929--1937}.
\newblock
\href{https://doi.org/10.1145/3534678.3539382}{doi:\nolinkurl{10.1145/3534678.3539382}}


\bibitem[Xi et~al\mbox{.}(2024)]%
        {Xi2024MemoCRS}
\bibfield{author}{\bibinfo{person}{Yunjia Xi} {et~al\mbox{.}}} \bibinfo{year}{2024}\natexlab{}.
\newblock \showarticletitle{{MemoCRS}: Memory-Enhanced Sequential Conversational Recommender Systems with Large Language Models}. In \bibinfo{booktitle}{\emph{Proceedings of the 33rd {ACM} International Conference on Information and Knowledge Management}}. \bibinfo{publisher}{ACM}, \bibinfo{address}{Boise, Idaho, USA}, \bibinfo{numpages}{10}~pages.
\newblock
\href{https://doi.org/10.1145/3627673.3679599}{doi:\nolinkurl{10.1145/3627673.3679599}}


\bibitem[Yang et~al\mbox{.}(2024)]%
        {Yang2024BehaviorAlignment}
\bibfield{author}{\bibinfo{person}{Dayu Yang}, \bibinfo{person}{Fumian Chen}, {and} \bibinfo{person}{Hui Fang}.} \bibinfo{year}{2024}\natexlab{}.
\newblock \showarticletitle{Behavior Alignment: A New Perspective of Evaluating {LLM}-based Conversational Recommender Systems}. In \bibinfo{booktitle}{\emph{Proceedings of the 47th International {ACM} {SIGIR} Conference on Research and Development in Information Retrieval}}. \bibinfo{publisher}{ACM}, \bibinfo{address}{Washington, DC, USA}, \bibinfo{pages}{1--5}.
\newblock
\urldef\tempurl%
\url{https://arxiv.org/abs/2404.11773}
\showURL{%
\tempurl}


\bibitem[Yang and Chen(2024)]%
        {Yang2024ReFICR}
\bibfield{author}{\bibinfo{person}{Ting Yang} {and} \bibinfo{person}{Li Chen}.} \bibinfo{year}{2024}\natexlab{}.
\newblock \showarticletitle{Unleashing the Retrieval Potential of Large Language Models in Conversational Recommender Systems}. In \bibinfo{booktitle}{\emph{Proceedings of the 18th {ACM} Conference on Recommender Systems}}. \bibinfo{publisher}{ACM}, \bibinfo{address}{Bari, Italy}, \bibinfo{pages}{1--10}.
\newblock
\href{https://doi.org/10.1145/3640457.3688146}{doi:\nolinkurl{10.1145/3640457.3688146}}


\bibitem[Yao et~al\mbox{.}(2024)]%
        {Yao2024TauBench}
\bibfield{author}{\bibinfo{person}{Shunyu Yao}, \bibinfo{person}{Noah Shinn}, \bibinfo{person}{Pedram Razavi}, {and} \bibinfo{person}{Karthik Narasimhan}.} \bibinfo{year}{2024}\natexlab{}.
\newblock \bibinfo{title}{$\tau$-Bench: A Benchmark for Tool-Agent-User Interaction in Real-World Domains}.
\newblock
\showeprint[arxiv]{2406.12045}~[cs.AI]
\urldef\tempurl%
\url{https://arxiv.org/abs/2406.12045}
\showURL{%
\tempurl}


\bibitem[Zhang et~al\mbox{.}(2016)]%
        {Zhang2016CKE}
\bibfield{author}{\bibinfo{person}{Fuzheng Zhang}, \bibinfo{person}{Nicholas~Jing Yuan}, \bibinfo{person}{Defu Lian}, \bibinfo{person}{Xing Xie}, {and} \bibinfo{person}{Wei-Ying Ma}.} \bibinfo{year}{2016}\natexlab{}.
\newblock \showarticletitle{Collaborative Knowledge Base Embedding for Recommender Systems}. In \bibinfo{booktitle}{\emph{Proceedings of the 22nd {ACM} {SIGKDD} International Conference on Knowledge Discovery and Data Mining}}. \bibinfo{publisher}{ACM}, \bibinfo{address}{San Francisco, California, USA}, \bibinfo{pages}{353--362}.
\newblock
\href{https://doi.org/10.1145/2939672.2939673}{doi:\nolinkurl{10.1145/2939672.2939673}}


\bibitem[Zhou et~al\mbox{.}(2023)]%
        {Zhou2023WebArena}
\bibfield{author}{\bibinfo{person}{Shuyan Zhou}, \bibinfo{person}{Frank~F. Xu}, \bibinfo{person}{Hao Zhu}, \bibinfo{person}{Xuhui Zhou}, \bibinfo{person}{Robert Lo}, \bibinfo{person}{Abishek Sridhar}, \bibinfo{person}{Xianyi Cheng}, \bibinfo{person}{Tianyue Ou}, \bibinfo{person}{Yonatan Bisk}, \bibinfo{person}{Daniel Fried}, \bibinfo{person}{Uri Alon}, {and} \bibinfo{person}{Graham Neubig}.} \bibinfo{year}{2023}\natexlab{}.
\newblock \bibinfo{title}{{WebArena}: A Realistic Web Environment for Building Autonomous Agents}.
\newblock
\showeprint[arxiv]{2307.13854}~[cs.AI]
\urldef\tempurl%
\url{https://arxiv.org/abs/2307.13854}
\showURL{%
\tempurl}


\bibitem[Zou et~al\mbox{.}(2024)]%
        {zou2024knowledge}
\bibfield{author}{\bibinfo{person}{Jie Zou}, \bibinfo{person}{Aixin Sun}, \bibinfo{person}{Cheng Long}, {and} \bibinfo{person}{Evangelos Kanoulas}.} \bibinfo{year}{2024}\natexlab{}.
\newblock \showarticletitle{Knowledge-Enhanced Conversational Recommendation via Transformer-Based Sequential Modeling}.
\newblock \bibinfo{journal}{\emph{{ACM} Transactions on Information Systems}} \bibinfo{volume}{42}, \bibinfo{number}{6} (\bibinfo{year}{2024}), \bibinfo{pages}{1--27}.
\newblock


\end{thebibliography}

\end{document}